\documentclass[aps,prb,10pt,notitlepage,twocolumn]{revtex4}

\usepackage{graphicx}
\usepackage{amsmath}
\usepackage{amssymb}
\usepackage{amsfonts}
\usepackage{color}

\newcommand{\pdmit}{Pd(dmit)$_2$}
\newcommand{\BCS}{|\text{BCS}\rangle}
\newcommand{\SP}{|\text{SP}\rangle}
\newcommand{\pBCS}{|\Psi_{\text{BCS}}\rangle}
\newcommand{\pSP}{|\Psi_{\text{SP}}\rangle}

\begin{document}

\title{Importance of anisotropy in the spin-liquid candidate Me$_{3}$EtSb[Pd(dmit)$_2$]$_2$ }

\author{A. C. Jacko}
\author{Luca F. Tocchio}
\author{Harald O. Jeschke}
\author{Roser Valent\'\i} 
\affiliation{Institut f\"ur Theoretische Physik, Goethe-Universit\"at Frankfurt, Max-von-Laue-Stra{\ss}e 1, 60438 Frankfurt am Main, Germany}

\begin{abstract}
  Organic charge transfer salts based on the molecule {\pdmit} display
  strong electronic correlations and geometrical frustration, leading
  to spin liquid, valence bond solid, and superconducting states,
  amongst other interesting phases. The low energy electronic degrees
  of freedom of these materials are often described by a single band
  model; a triangular lattice with a molecular orbital representing a
  {\pdmit} dimer on each site. We use \textit{ab initio} electronic
  structure calculations to construct and parametrize low energy
  effective model Hamiltonians for a class of Me$_{4-n}$
  Et$_nX$[Pd(dmit)$_2$]$_2$ ($X$=As, P, N, Sb) salts and investigate how
  best to model these systems by using variational Monte Carlo (VMC)
  simulations. Our findings suggest that the prevailing model of these
  systems as a $t-t'$ triangular lattice is incomplete, and that a
  fully anisotropic triangular lattice (FATL) description produces
  importantly different results, including a significant lowering of
  the critical $U$ of the spin-liquid phase.
\end{abstract}

\maketitle

\section{Introduction} 

The Me$_{4-n}$Et$_nX$[Pd(dmit)$_2$]$_2$ family
of organic crystals~\cite{chemistry} is known for its many interesting electronic and
magnetic phases; these materials can have superconducting, Mott
insulating, spin-liquid, valence bond solid and spin density wave
orders, determined by the cation as well as by temperature and applied
pressure.\cite{kobayashi91,seya95,tamura02,itou08} The reason for this
rich variety of phases is the competition between frustration effects
and electronic correlations. As such, these materials are the focus of
very active research. Here we will investigate the significance of
anisotropy and correlations in these materials by parameterizing and
solving model Hamiltonians. We will abbreviate Me$_{4-n}$Et$_n
X$[Pd(dmit)$_2$]$_2$ as $X$-$n$, following Ref.~\onlinecite{powell11}.

\begin{figure}
\begin{center}
\includegraphics[width=0.8\columnwidth]{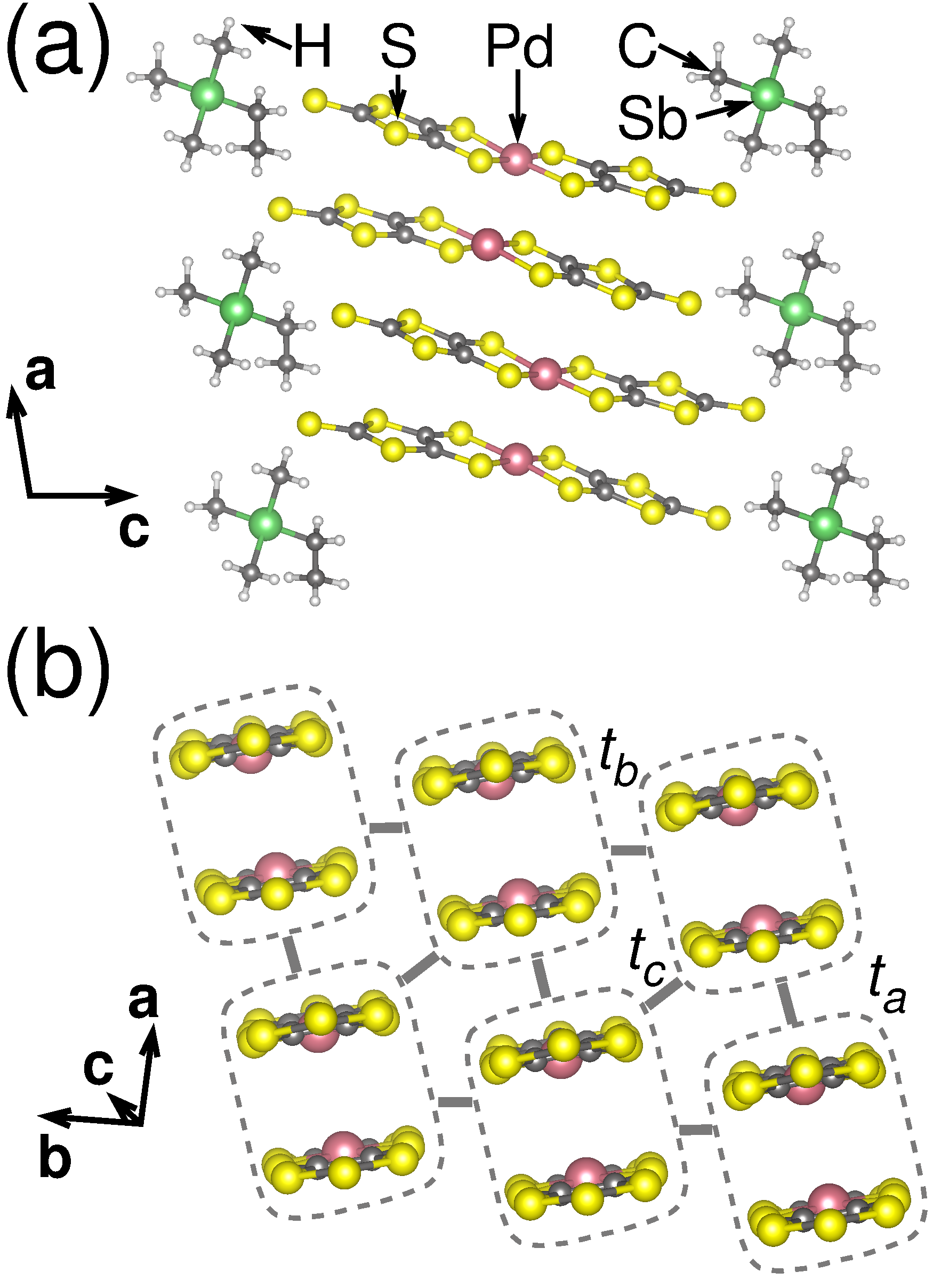}  	
\caption{(a) Structure of the spin-liquid candidate
  Me$_3$EtSb[Pd(dmit)$_2$]$_2$ (Sb-1). (b) Dimerization of the
  {\pdmit} molecules, where dimers form 2D triangular lattice layers
  in the $a$-$b$ plane, separated along $c$ by cation layers (in this
  case, Me$_3$EtSb). We follow Ref.~\protect\onlinecite{nakamura12} in labeling the
  hopping integrals in the $a$-$b$ plane $t_a$, $t_b$ and
  $t_c$.}\label{fig:sb1acplane}
\end{center}
\end{figure}

These materials share a generic crystal structure, illustrated in
Fig.~\ref{fig:sb1acplane}, and are found in a variety of
ordered phases.

As-0 enters an anti-ferromagnetic (AFM) phase below 35 K; Sb-1 has a
spin-liquid ground state, and spin susceptibility measurements imply
that it has an exchange interaction $220\,{\rm K}  \leq J \leq 250\,{\rm K}$; Sb-2
has no low temperature AFM transition (unlike P-2 and most of the
$X$-0s); N-0 is the only $X$-0 material with a superconducting transition
(6.2 K at 6.5 kbar), while at lower pressures (away from the superconducting phase)
it exhibits a spin-density wave
transition.\cite{kobayashi91,seya95,tamura02,itou08} Developing a
unified description of this wide variety of phases is an ongoing
challenge.

Here we calculate the electronic structure and use Wannier orbitals
for the frontier bands to parameterize model Hubbard Hamiltonians.  By
solving these Hamiltonians with variational Monte Carlo (VMC), we show
that there are important qualitative differences between treating the
system as a $t-t'$ triangular lattice and considering it as a fully
anisotropic triangular lattice (FATL). A discussion on the role of
full anisotropy within the Heisenberg model may be found in
Ref.~\onlinecite{hauke13}.

\section{Methods}

We perform density functional theory calculations of the electronic
structure and then construct localized Wannier orbitals for the
frontier bands. From these we parameterize tight-binding model
Hamiltonians, both for the $t-t'$ and for the FATL ($t_a-t_b-t_c$).
We then solve these models with a VMC approach. The approach used here
allows for the description of metallic and spin-liquid states with the
same variational wavefunction, which is compared to a variational
wavefunction that describes the magnetic spiral ordered insulating
state.

\subsection{Electronic Structure}  

The electronic structure calculations
presented here were performed in an all-electron full-potential local
orbital basis using the FPLO package.\cite{koepernik99} The densities
were converged on a $(6 \times 6\times 6)$ $k$ mesh using a
generalized gradient approximation (GGA) functional.\cite{perdew96}

To move from density functional theory calculations to model
Hamiltonians, it is convenient to construct Wannier orbitals to
represent the frontier bands of the system. In principle the Wannier
orbitals are simply Fourier transforms of the Bloch wavefunctions;
however in this procedure there are still many degrees of
freedom. Here they are constrained by the requirements that the
Wannier orbitals be represented by Kohn-Sham orbitals in a narrow
energy window; by projecting onto the FPLO basis orbitals ensures that
they will form a good basis for a tight-binding model.\cite{eschrig09}
With these nearly optimally localized Wannier orbitals we compute
real-space overlaps to obtain tight-binding parameters.  This method
has several advantages over band fitting, as has been discussed
previously in the case of molecular organic crystals.\cite{jacko13a}

\subsection{Effective Modelling}  

To model these systems we consider the
Hubbard Hamiltonian. As we will show later, a half-filled single
orbital per site Hubbard model is suggested by the electronic
structure, so it is this kind of model we will focus on. The
Hamiltonian is defined by:
\begin{equation}\label{eq:hubbard}
{\cal H}=-\sum_{i,j,\sigma} t_{ij} c^\dagger_{i,\sigma} c_{j,\sigma} 
+ U \sum_{i} n_{i,\uparrow} n_{i,\downarrow},
\end{equation}
where $c^\dagger_{i,\sigma} (c_{i,\sigma})$ creates (destroys) an electron 
with spin $\sigma$ on site $i$, $n_{i,\sigma}=c^\dagger_{i,\sigma}c_{i,\sigma}$
is the electronic density, $t_{ij}$ is the hopping amplitude and $U$ is the 
on-site Coulomb repulsion. Here we calculate a VMC solution including backflow correlations, 
which allows us to provide an accurate description of a system with hundreds of lattice sites.\cite{tocchio11}

A major issue in the Hubbard model on the anisotropic triangular lattice 
is the possibility of stabilizing a spin-liquid phase (as seen in the 
experimental data). In addition, for generic values of the hopping parameters magnetic 
states with generic spiral order may be expected and indeed can be obtained 
within mean-field approaches, like for instance the Hartree-Fock
(HF) approximation.~\cite{weihong99,krishnamurthy90}

Here, we approach this problem by implementing correlated variational wave 
functions which describe both spin-liquid states and magnetic states with generic ordering vectors. 
In this way, we are 
able to treat spiral order and paramagnetic  
states at the same level of theory, and therefore have 
a sensible comparison of their energies vs $U$.\cite{tocchio13}

\textit{Spiral Magnetic States.-}
We start with the magnetic states obtained at the HF level, with the only constraint for the spin order 
to be coplanar in the $x-y$ plane. 
Our magnetic HF solutions display spiral order, 
which (in 2D) may be parametrized by two pitch 
angles $\theta$ and $\theta'$, where $\theta$ ($\theta'$) is the angle between 
nearest-neighbor spins along the hopping direction $t_b$ ($t_c$).
Since we use finite clusters, only certain commensurate angles are allowed.
For a lattice size $L=l \times l$, the allowed values are $\theta=2\pi n/l$ and $\theta'=2\pi n'/l$, 
with $n$ and $n'$ integers. We tested various lattice sizes ranging from $12\times 12$ to $20\times 20$, 
and determined the best pair of pitch angles for each lattice size.

Our VMC magnetic states are then constructed by applying correlation terms on top 
of the HF spiral states $\SP$. We employ a spin-spin Jastrow factor to correctly describe
fluctuations orthogonal to the plane where the magnetic order lies, i.e., 
${\cal J}_s=\exp [\frac{1}{2} \sum_{i,j} u_{i,j} S_i^z S_j^z ]$.~\cite{franjic97}
A further density-density Jastrow factor
${\cal J}_c=\exp [\frac{1}{2} \sum_{i,j} v_{i,j} n_i n_j ]$ 
(that includes the on-site Gutzwiller term $v_{i,i}$) is considered to adjust 
electron correlations. All the $u_{i,j}$'s and the $v_{i,j}$'s are optimized 
for each independent distance $|i-j|$. The correlated state is then given by 
$|\Psi_{\textrm{SP}}\rangle = {\cal J}_s {\cal J}_c \SP$. 

\textit{Paramagnetic States.-}
In order to describe a paramagnetic state, we construct 
an uncorrelated wave function given by the ground state $\BCS$
of a superconducting BCS Hamiltonian:~\cite{zhang88,edegger07}
\begin{equation}\label{eq:meanfield}
\begin{split}
{\cal H}_{\rm{BCS}} = & \sum_{i,j,\sigma} \tilde{t}_{ij} 
c^\dagger_{i,\sigma} c_{j,\sigma} -\mu \sum_{i,\sigma} c^\dagger_{i,\sigma} c_{i,\sigma} \\
& + \sum_{i,j} \Delta_{ij} 
c^\dagger_{i,\uparrow} c^{\dagger}_{j,\downarrow} + \textrm{h.c.},
\end{split}
\end{equation}
where both the variational hopping amplitudes $\tilde{t}_{ij}$, the pairing
fields $\Delta_{ij}$, and the chemical potential $\mu$ are variational
parameters to be independently optimized.  For the majority of the
results reported here we constrain all of these variational parameters
to be real.

The correlated state $\pBCS={\cal J}_c \BCS$ allows us to describe a
paramagnetic Mott insulator for a sufficiently singular Jastrow factor
$v_q\sim 1/q^2$ ($v_q$ being the Fourier transform of
$v_{i,j}$),~\cite{capello05} while a metallic state can be obtained
whenever $v_q\sim 1/q$.

A size-consistent and efficient way to further improve the correlated
states $\pBCS$ and $\pSP$ is based on backflow correlations. In this
approach, each orbital that defines the unprojected states $\BCS$ and
$\SP$ is taken to depend upon the many-body configuration, in order to
incorporate virtual hopping processes.~\cite{tocchio11} All results
presented here are obtained by fully incorporating the backflow
corrections and optimizing individually every variational parameter in
the mean-field BCS equation, and in the Jastrow factors ${\cal J}_c$
and ${\cal J}_s$, as well as in the backflow corrections.

\section{Determination of an appropriate model} 

The spin-liquid
candidate Sb-1 has been the subject of much recent study. We will
focus on this material in discussing the process of determining the
appropriate model Hamiltonian for this class of systems.  To move
towards the goal of understanding the phase diagram of Sb-1, and the
origin of the various ordered phases, one needs a sensible choice of a
model Hamiltonian with reliable parameters. In constructing some
minimal model, one must also be aware of what the model neglects, and
what effect that has on the physics it predicts. Here we analyze the
electronic structure to determine an appropriate minimal model, and
then use Wannier orbitals to parameterize it.

\begin{figure}
\begin{center}
 \includegraphics[width=0.9\columnwidth]{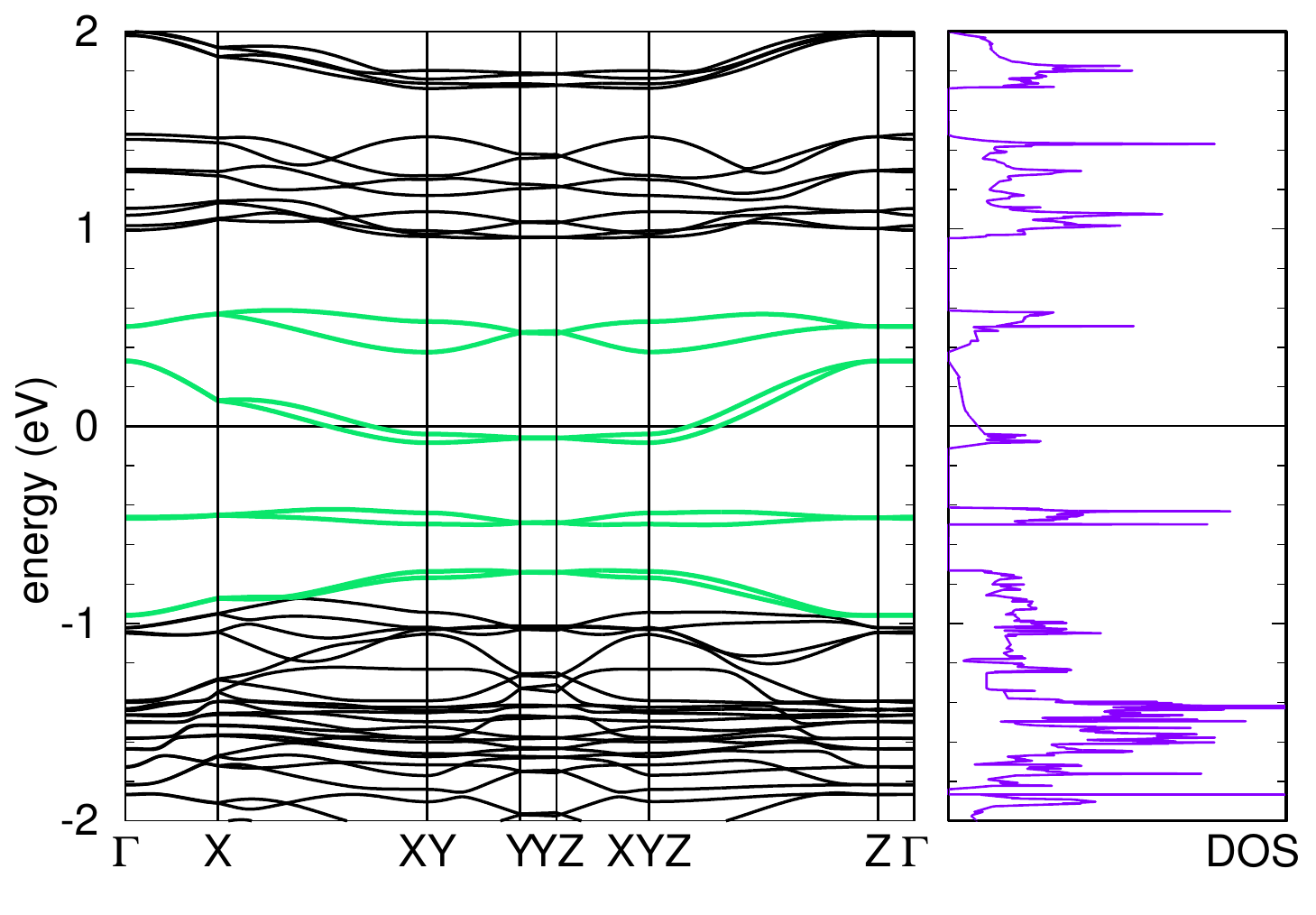} 
 \caption{Band structure and density of states of Sb-1 in a wide
   energy window around the Fermi energy. The four pairs of bands
   highlighted in green are those identified as being of {\pdmit} HOMO
   or LUMO origin. The pair that crosses the Fermi energy is of
   HOMO-ab origin.  These are the bands used in a minimal one-orbital
   model of the system. The direct energy gap between the HOMO-ab
   bands and the others is more than 0.1 eV, although the indirect gap
   is smaller. }\label{fig:bs_sb1}
\end{center}
\end{figure}

\begin{figure}
\begin{center}
 \includegraphics[width=0.6\columnwidth]{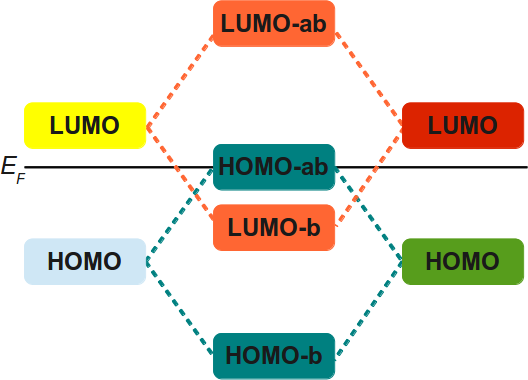} 
 \caption{Schematic of the hybridization of the orbital of the two
   {\pdmit} in each dimer, and the resultant crossing of energy levels
   leading to the unusually ordered frontier bands of {\pdmit}
   complexes.}\label{fig:energylevel}
\end{center}
\end{figure}

Fig.~\ref{fig:bs_sb1} shows the band structure and the density of
states for Sb-1 in a wide energy window around the Fermi energy.  From
this figure, we can see that there are several pairs of bands
separated from the others; these bands are the bands identified as
arising from bonding (-b) and anti-bonding (-ab) hybrids of {\pdmit}
highest occupied molecular orbitals (HOMO) and lowest unoccupied
molecular orbitals (LUMO) (highlighted in green).  The pair crossing
the Fermi energy are HOMO-ab bands, while the pairs on either side are
the LUMO-b and -ab bands.  The fourth highlighted pair at the top of
the bulk valence bands is the HOMO-b pair. The origin of these bands
is illustrated in Fig.~\ref{fig:energylevel}. The bands come in pairs
because for each dimer orbital in one plane, there is an identical one
in the other plane, related to the first by a translation and a
rotation about the $t_b$ direction (the $y$ axis).  These pairs of
bands are well separated from each other, with typical direct band
gaps on the order of 100's of meV, well above $k_B T$ for most
experiments on these systems.

\subsection{Modeling the Frontier Bands}  

\begin{figure}
\begin{center}
 \includegraphics[width=\columnwidth]{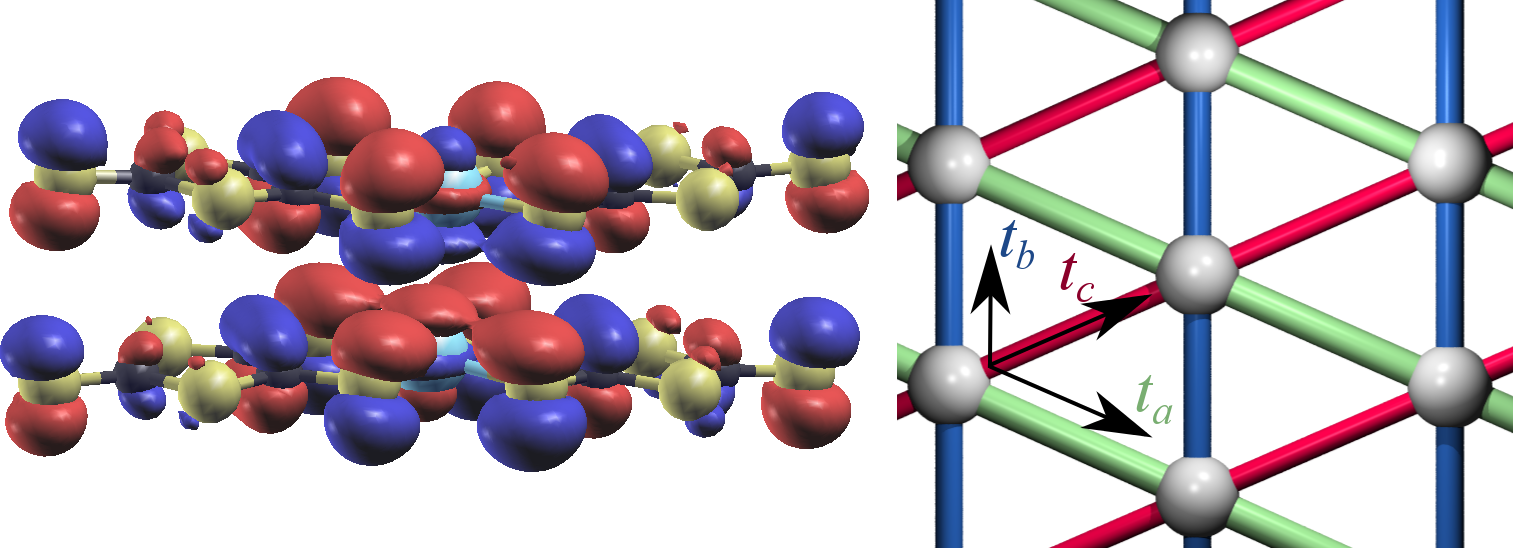} 	
 \caption{Left panel: Wannier orbital for the HOMO-ab bands of Sb-1. The anti-bonding HOMO character of the orbital is clearly
   visible. Right panel: The 2D tight-binding lattice generated in the $a$-$b$ plane from the Wannier orbital of the left panel. 
   Each grey point of the lattice represents a dimer. The widths of the various cylinders in the lattice are
   linearly proportional to the magnitude of the corresponding $t$
   (see Table~\ref{tab:allts}).}\label{fig:WF_TBH_Sb1}
\end{center}
\end{figure}

As Fig.~\ref{fig:bs_sb1} shows,
the HOMO-ab bands are well separated from the other bands, and so they
form a good basis for a low energy effective model Hamiltonian. In
Fig.~\ref{fig:WF_TBH_Sb1} we show one of the Wannier orbitals for
these bands, clearly showing their HOMO-ab character (the other
Wannier orbital is the same but in the other layer of {\pdmit}
molecules).  Table~\ref{tab:allts} shows the three $t$'s in the
first-nearest neighbour shell (see Fig.~\ref{fig:WF_TBH_Sb1}) computed
from the HOMO-ab Wannier orbitals of several $X$-$n$ systems (note
that including more neighbours explicitly has no effect on these
parameters, unlike in a band-fitting computation).  These results are
consistent with those found in past
work.\cite{scriven12,nakamura12,tsumuraya13}

\begin{table}
\begin{tabular}{|l|cccc|c|}
\hline
$X$-$n$ 		& 	$\mu$ 	& 	$t_{b}$ & 	$t_{a}$ & 	$t_{c}$  & Ref. \\ \hline
N-0 		&	34.8	&	44.3	&	48.6	&	38.9 & \onlinecite{kobayashi90}*\\
As-0 	 	&	28.6	&	44.5	&	55.6	&	32.6 & \onlinecite{kobayashi90}*\\ 
P-1 		&	29.3	&	39.8	&	48.4	&	46.4 & \onlinecite{kato06} \\ 
Sb-1 		&	32.3	&	46.9	&	56.5	&	39.8 & \onlinecite{katosb1} \\
Sb-2		&	32.4	&	35.2	&	45.5	&	44.1 & \onlinecite{nakao05}* \\ \hline
\end{tabular}
\caption{Comparison of the sets of one orbital model parameters for several $X$-$n$ systems. 
  All energies are given in meV. Starred references did not include H coordinates so they were inserted manually. 
  All structures used here were obtained at room temperature, except for Sb-1 which was obtained at 4K.}\label{tab:allts}
\end{table}

\begin{figure}
\includegraphics[width=\columnwidth]{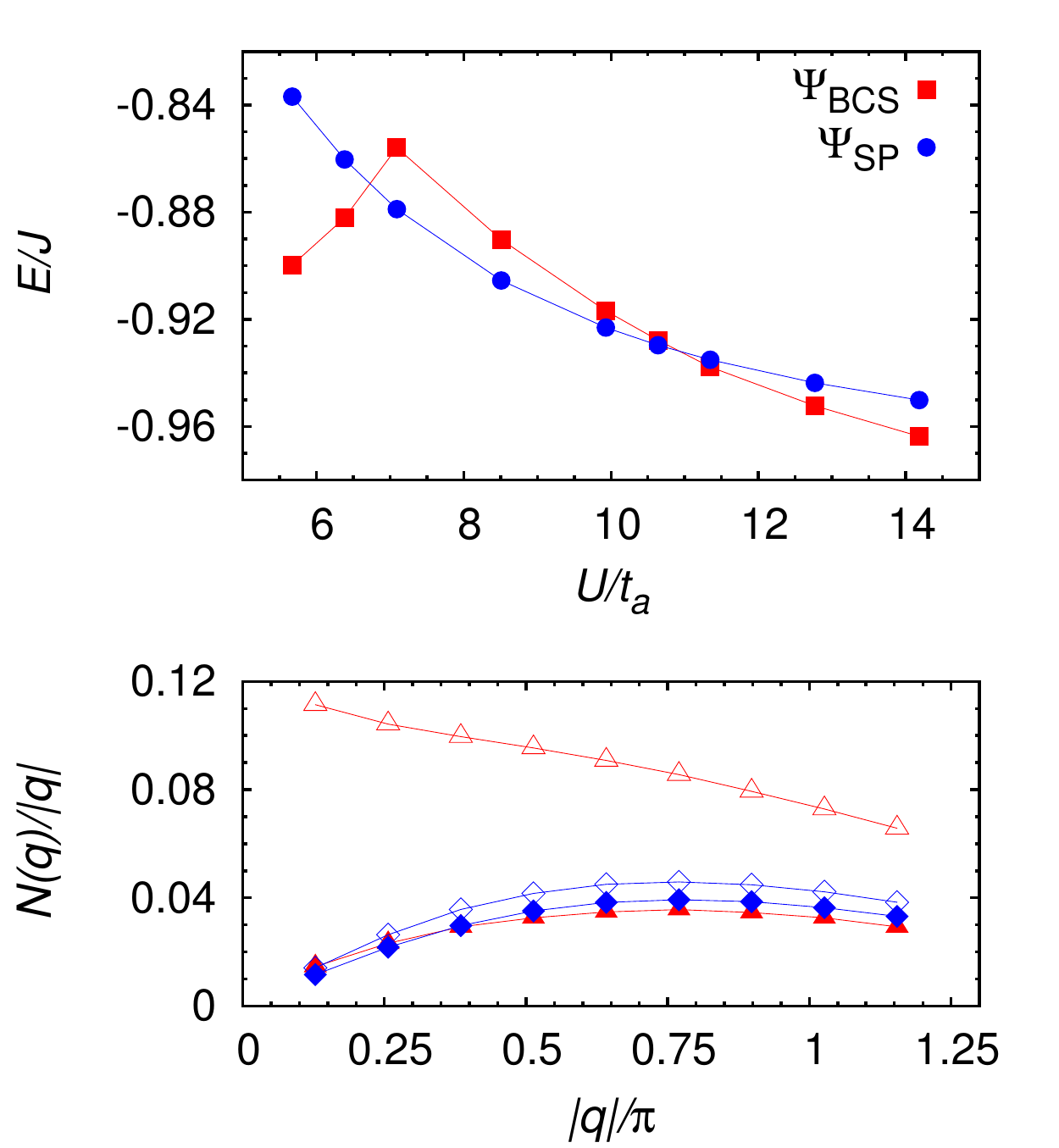}
\caption{\label{fig:wfnenergygap} (Color online) Upper panel:
  Variational energies per site for the FATL model with the parameters
  for Sb-1, by using the paramagnetic state (red squares), $\pBCS$,
  and the magnetic spiral state (blue circles), $\pSP$, as a function
  of $U/t_a$ and in units of $J=4t_a^2/U$.  Both wavefunctions were
  computed on a $L=324$ lattice, for which the optimal pitch angles
  ($\theta = 7\pi/9, \theta'=3\pi/9$) are commensurate.  Lower panel:
  $N(q)/q$ at $U/t_a = 6.4$ by using $\pBCS$ (empty triangles) and
  $\pSP$ (empty diamonds) and at $U/t_a = 7.1$ by using $\pBCS$ (full
  triangles) and $\pSP$ (full diamonds). Data show the metal to
  insulator transition in $\pBCS$, while the magnetic state is always
  insulating.  Plots are presented along the line connecting the point
  $Q=(\pi,\pi/\sqrt{3})$ to the point $\Gamma=(0,0)$ in reciprocal
  space.}
\end{figure}

With the $t$'s obtained from the Wannier orbitals, we can now explore
the Hubbard model for Sb-1 with VMC.  We find that the spiral
magnetic state $\SP$ has optimal pitch angles $\theta = 7\pi/9,
\theta'=3\pi/9$, that are commensurate to an $18\times 18$ lattice size.  The
BCS wavefunction has finite pairing fields for $U/t_a > 6.75$
(\textit{i.e.} when the system is insulating) and they are highly
anisotropic; with the largest component along the $t_a$ direction, the
$t_b$ component approximately half as large and with opposite sign,
and the $t_c$ component nearly zero.  Fig.~\ref{fig:wfnenergygap}
shows the optimized energies of these two wavefunctions as a function
of $U/t_a$; $\pSP$ is favorable for $6.75 < U/t_a < 11$, and $\pBCS$
is favorable outside of this region.

The charge gap, $G$, can be calculated from the static structure
factor, $N(q)$, by assuming that the low momentum excitations are
collective modes.\cite{feynman54,tocchio11} With this approximation,
one finds that $G \propto \lim_{q\to 0}\frac{q^2}{N(q)}$, where
$N(q)=\langle n_{-q}n_q\rangle$ and $n_q=1/\sqrt{L}\sum_{r,\sigma}
e^{iqr}n_{r,\sigma}$. As such, the metallic phase is characterized by
$N(q)/q \to const.$ as $q\to 0$, implying a vanishing gap at $q=0$,
and the insulating phase by $N(q)/q \to 0$ as $q \to 0$, implying a
finite gap. Fig.~\ref{fig:wfnenergygap} shows $N(q)/q$ versus $q$ for
the $\pBCS$ and $\pSP$ states at $U/t_a=6.4$ and $U/t_a=7.1$, either
side of the point the $\pSP$ state becomes favorable. While the spiral
magnetic $\pSP$ state is insulating on both sides of the transition,
the $\pBCS$ state changes from metallic to insulating. Thus, we
clearly see this is the metal-insulator transition (MIT).  By
examining $N(q)/q$, we confirm that the $\pBCS$ state remains
insulating all the way above the MIT, in particular in the region
$U/t_a > 11$, where it becomes favorable.

We note that an existing calculation of the interaction parameters
using the constrained random-phase approximation (cRPA) finds $U/t_a
\sim 11$ for Sb-1, in good agreement with our results for the location
of the spin-liquid region.\cite{nakamura12}

\subsection{$t-t'$ model VS FATL}  

\begin{figure}
\begin{center}
 \includegraphics[width=0.9\columnwidth]{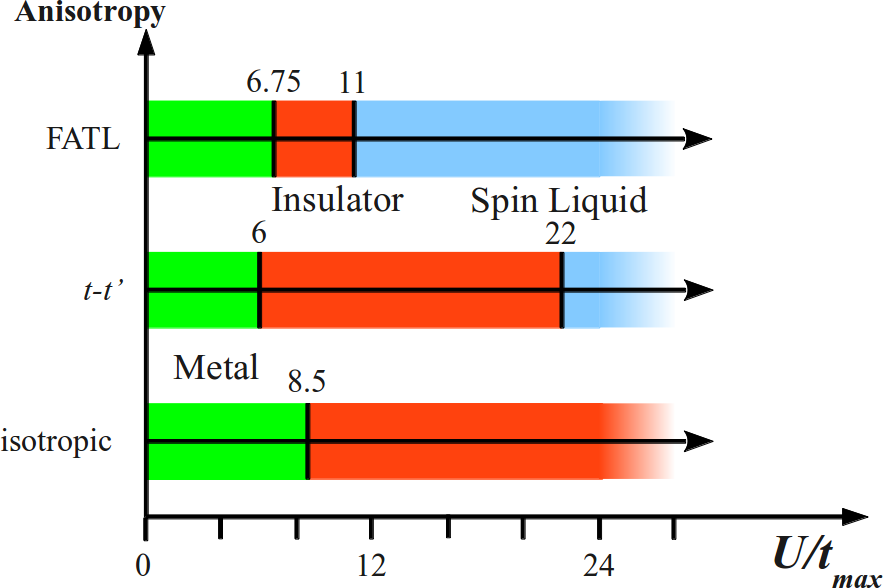} 	
 \caption{Phase diagram versus $U/t_{\textrm{max}}$, where
   $t_{\textrm{max}}$ is the biggest hopping parameter in the model,
   for the isotropic, $t-t'$, and FATL models for Sb-1; showing the
   metallic (green), spiral ordered magnetic insulator (orange) and
   spin liquid (cyan) phases.  The spin-liquid phase is strongly
   enhanced by including the full anisotropy of the system, while the
   metal-insulator transition is only slightly
   changed.}\label{fig:phasediagram}
\end{center}
\end{figure}

{\pdmit} systems are often represented
by a $t-t'$ model, despite this symmetry not being found in the
various $t$ estimates, including our estimate in
Table~\ref{tab:allts}.\cite{scriven12,nakamura12,tsumuraya13} Here we
compare model results using the $t-t'$ approximation with a fully
anisotropic triangular lattice (FATL) model, focusing our discussion
on the spin-liquid candidate Sb-1.  If we average the two larger $t$'s
of Sb-1, assuming $t=(t_a+t_b)/2$ and $t'=t_c$, we find that the
equivalent $t-t'$ model has $t'/t = 0.77$.  This model has been
previously studied with variational Monte Carlo,\cite{tocchio13} where
for this value of $t'/t$ the critical $U$ for the spin-liquid
transition is located at $U/t\sim 22$, while for the FATL this value
is strongly reduced to $U/t_a\sim 11$. On the other hand, the
metal-insulator transition is only slightly affected, raising from
$U_c/t \sim 6$ for the $t-t'$ model to $U_c/t_a\sim 6.75$ for the
FATL.  These results are illustrated in the phase diagram in
Fig.~\ref{fig:phasediagram}.

We would like to point out that if we repeat our calculation for the
P-1 system (which is almost truly $t-t'$) we find that in both the
FATL and in the $t-t'$ model, the transition to spin liquid occurs at
$U/t_{\textrm{max}}\sim 13$, (where $t_{\textrm{max}}$ is the biggest
hopping parameter). Thus the enhancement of the spin-liquid phase is
driven by anisotropy.

Interestingly, if we allow the pairing fields to be complex, we find
that finite imaginary components lower the energy of the $\pBCS$ state
slightly.  This imaginary component appears for $U/t_a > 11$, the same
point where the $\pBCS$ state becomes favorable again and it seems to
be due to the anisotropy of the system; indeed by considering the less
anisotropic P-1 system, we find a smaller imaginary component, while
it is not seen in $t-t'$ models.

\section{Conclusions} 

We have shown that the high anisotropy in
{\pdmit} materials can have important effects on the physics.
Our electronic structure calculations find a range of different hopping
parameters in good agreement with previous DFT calculations.
Relative to a $t-t'$, using a FATL model for Sb-1 increases the
critical $U$ for the metal-insulator transition only by $\sim 10\%$,
while it \emph{halves} the critical $U$ for the spin-liquid
transition.  With this reduction, existing parameterisations of Sb-1
place it in the spin-liquid regime.  In addition, the spin-liquid
phase develops a complex pairing function as it becomes favorable for
$U/t_a > 11$, something not seen in the $t-t'$ model.

\acknowledgments

We would like to acknowledge D. Cocks,
C. Gros, M. Imada, and H. Seo for useful discussions and H. Feldner for providing the HF code.  
We acknowledge the support of the German
Science Foundation through the grant SFB/TR49.

%

\end{document}